\documentclass[aps,floatfix,nofootinbib,showpacs,twocolumn,superscriptaddress]{revtex4}

\usepackage{amsmath}
\usepackage{amssymb}
\usepackage{amsthm}
\usepackage{dcolumn}
\usepackage{epsfig}
\usepackage{graphics}
\usepackage{graphicx}
\usepackage{amsmath}
\usepackage{longtable}
\usepackage{color}

\definecolor{purple}{rgb}{0.5,0,0.5}
\definecolor{blue}{rgb}{0.0,0,0.9}
\usepackage[colorlinks=true, pdfstartview=FitV, linkcolor=purple, citecolor= purple, urlcolor=blue]{hyperref}

\newcommand{\bea}{\begin{eqnarray}}
\newcommand{\eea}{\end{eqnarray}}

\newcommand{\nn}{\nonumber}

\newcommand{\beas}{\begin{eqnarray*}}
\newcommand{\eeas}{\end{eqnarray*}}

\begin{document}
\title{Pion form factor from a contact interaction}

\author{L.\,X.~Guti\'errez-Guerrero}
\affiliation{Instituto de F{\i}sica y Matem\'aticas,
Universidad Michoacana de San Nicol\'as de Hidalgo, Apartado Postal
2-82, Morelia, Michoac\'an 58040, Mexico}

\author{A.~Bashir}
\affiliation{Instituto de F{\i}sica y Matem\'aticas,
Universidad Michoacana de San Nicol\'as de Hidalgo, Apartado Postal
2-82, Morelia, Michoac\'an 58040, Mexico}

\author{I.\,C.~Clo\"et}
\affiliation{Department of Physics, University of Washington, Seattle WA 98195, USA}

\author{C.\,D.~Roberts}
\affiliation{Physics Division, Argonne National Laboratory, Argonne, Illinois 60439, USA}
\affiliation{Department of Physics, Peking University, Beijing 100871, China}

\begin{abstract}
In a Poincar\'e-covariant vector-boson-exchange theory, the pion possesses components of pseudovector origin, which materially influence its observable properties.  For a range of such quantities, we explore the consequences of a momentum-independent interaction, regularised in a symmetry-preserving manner.  The contact interaction, whilst capable of describing pion static properties, produces a form factor whose: evolution for $Q^2>0.17\,$GeV$^2$ disagrees markedly with experiment; and asymptotic power-law behaviour conflicts strongly with perturbative-QCD.
\end{abstract}


\maketitle

\date{\today}

%
The pion has a unique place in the Standard Model.  It is a bound-state of a dressed-quark and -antiquark, and also that almost-massless collective excitation which is the Goldstone mode arising from the dynamical breaking of chiral symmetry.  This dichotomy can only be understood by merging the study of many-body aspects of the QCD vacuum with the symmetry-preserving analysis of two-body bound-states.  Furthermore, the possibility that this dichotomous nature could have wide-ranging effects on pion properties has made the empirical investigation of these properties highly desirable, despite the difficulty in preparing a system that can act as a pion target and the concomitant complexities in the interpretation of the experiments; e.g., \cite{Volmer:2000ek,Horn:2006tm,Tadevosyan:2007yd,Wijesooriya:2005ir}.

A true understanding of the pion is achievable via QCD's Dyson-Schwinger equations (DSEs) \cite{Roberts:2007jh}.  Key to this is the existence of a nonperturbative symmetry-preserving truncation scheme \cite{Munczek:1994zz,Bender:1996bb,Chang:2009zb}, which enables: the connection to be made between dynamical chiral symmetry breaking (DCSB) and the bound-state pion \cite{Maris:1997hd}; and the use of experiment to chart QCD's $\beta$-function.  The $\beta$-function's perturbative evolution is well-known but this new feedback between experiment and nonperturbative theory may provide access to its long-range behaviour.

Studying the electromagnetic pion form factor, $F^{\rm em}_{\pi}(Q^2)$, is an ideal way to elucidate the potential of this interaction.  For example, there is a prediction \cite{Farrar:1979aw,Efremov:1979qk,Lepage:1980fj} that $Q^2 F_{\pi}(Q^2)\approx\,$constant for $Q^2\gg m_\pi^2$ in a theory whose interaction is mediated by massless vector-bosons.  The verification of this prediction is a strong motivation for modern experiment \cite{Volmer:2000ek,Horn:2006tm,Tadevosyan:2007yd}.  However, one may adopt a different view of this programme; namely, as an attempt to constrain and map experimentally the nature of the exchange interaction that binds the pion.

Poincar\'e covariance entails that the Bethe-Salpeter amplitude for an isovector pseudoscalar bound-state of a dressed-quark and -antiquark takes the form
\begin{eqnarray}
\nonumber
\lefteqn{\Gamma_{\pi}^j(k;P) = \tau^{j}\gamma_5 \left[ i E_\pi(k;P) + \gamma\cdot P F_\pi(k;P) \right.}\\
&& \left. + \,  \gamma\cdot k \, G_\pi(k;P) + \sigma_{\mu\nu} k_\mu P_\nu H_\pi(k;P) \right],
\label{genpibsa}
\end{eqnarray}
where $k$ is the relative and $P$ the total momentum of the constituents, and $\{\tau^j,j=1,2,3\}$ are the Pauli matrices.  (We employ a Euclidean metric with:  $\{\gamma_\mu,\gamma_\nu\} = 2\delta_{\mu\nu}$; $\gamma_\mu^\dagger = \gamma_\mu$; $\gamma_5= \gamma_4\gamma_1\gamma_2\gamma_3$; and $a \cdot b = \sum_{i=1}^4 a_i b_i$.)  This amplitude is determined from the homogeneous Bethe-Salpeter equation (BSE):
\begin{equation}
[\Gamma_{\pi}^j(k;P)]_{tu} = \int \! \frac{d^4q}{(2\pi)^4} [\chi_{\pi}^j(q;P)]_{sr} K_{tu}^{rs}(q,k;P)\,,
\label{genbse}
\end{equation}
where
%
%
$\chi_\pi^j(q;P) = S(q+P)\Gamma_{\pi}^j(q;P)S(q)$, $r,s,t,u$ represent colour, flavour and spinor indices, and $K$ is the quark-antiquark scattering kernel.  In Eq.\,(\ref{genbse}), $S$ is the dressed-quark propagator; viz., the solution of the gap equation:
\begin{eqnarray}
\nonumber \lefteqn{S(p)^{-1}= i\gamma\cdot p + m}\\
&&+ \int \! \frac{d^4q}{(2\pi)^4} g^2 D_{\mu\nu}(p-q) \frac{\lambda^a}{2}\gamma_\mu S(q) \frac{\lambda^a}{2}\Gamma_\nu(q,p) ,
\label{gendse}
\end{eqnarray}
wherein $m$ is the Lagrangian current-quark mass, $D_{\mu\nu}(k)$ is the gluon propagator and $\Gamma_\nu$ is the quark-gluon vertex.

QCD-based DSE calculations of $F^{\rm em}_\pi(Q^2)$ exist \cite{Maris:1998hc,Maris:2000sk}, the most systematic of which \cite{Maris:2000sk} predicted the measured form factor \cite{Volmer:2000ek}.  Our primary goal is to elucidate the sensitivity of $F^{\rm em}_\pi(Q^2)$ to the pointwise behaviour of the interaction between quarks.  We therefore describe how predictions for pion properties change if quarks interact instead through a contact interaction.

We thus begin by defining
\begin{equation}
\label{njlgluon}
g^2 D_{\mu \nu}(p-q) = \delta_{\mu \nu} \frac{1}{m_G^2}\,,
\end{equation}
where $m_G$ is a gluon mass-scale (such a scale is generated dynamically in QCD, with a value $\sim 0.5\,$GeV \cite{Bowman:2004jm}) and proceed by embedding this interaction in a rainbow-ladder truncation of the DSEs.  This means $\Gamma_{\nu}(p,q) =\gamma_{\nu}$ in both Eq.\,(\ref{gendse}) and the construction of $K$ in Eq. (\ref{genbse}).  Rainbow-ladder is the leading-order in a nonperturbative, symmetry-preserving truncation \cite{Munczek:1994zz,Bender:1996bb}.  It is known and understood to be accurate for pseudoscalar mesons \cite{Bhagwat:2004hn,Chang:2009zb} and guarantees current conservation \cite{Roberts:1994hh}.

With this interaction the gap equation becomes
\begin{equation}
 S^{-1}(p) =  i \gamma \cdot p + m +  \frac{4}{3}\frac{1}{m_G^2} \int\!\frac{d^4 q}{(2\pi)^4} \,
\gamma_{\mu} \, S(q) \, \gamma_{\mu}\,.   \label{gap-1}
\end{equation}
The integral possesses a quadratic divergence, even in the chiral limit.  If the divergence is regularised in a Poincar\'e covariant manner, then the solution is
\begin{equation}
S(p)^{-1} = i \gamma\cdot p + M\,,
\end{equation}
where $M$, momentum-independent, is determined by
\begin{equation}
M = m + \frac{M}{3\pi^2 m_G^2} \int_0^\infty \!ds \, s\, \frac{1}{s+M^2}\,.
\end{equation}

To proceed it is necessary to be specific about the regularisation procedure.  We write  \cite{Ebert:1996vx}
\begin{equation}
\frac{1}{s+M^2} = \int_0^\infty d\tau\,{\rm e}^{-\tau (s+M^2)}
\rightarrow \int_{\tau_{\rm uv}^2}^{\tau_{\rm ir}^2} d\tau\,{\rm e}^{-\tau (s+M^2)},
\end{equation}
where $\tau_{\rm ir,uv}$ are, respectively, infrared and ultraviolet regulators.  A nonzero value of $\tau_{\rm ir}=:1/\Lambda_{\rm ir}$ implements confinement by ensuring the absence of quark production thresholds \cite{Roberts:2007ji}.  Furthermore, since Eq.\,(\ref{njlgluon}) does not define a renormalisable theory,  $\Lambda_{\rm uv}:=1/\tau_{\rm uv}$ cannot be removed but instead plays a dynamical role and sets the scale of all dimensioned quantities.

The gap equation can now be written
\begin{equation}
M = m +  \frac{M}{3\pi^2 m_G^2} \,{\cal C}(M^2;\tau_{\rm ir},\tau_{\rm uv})\,,
\end{equation}
where ${\cal C}/M^2 = \Gamma(-1,M^2 \tau_{\rm uv}^2) - \Gamma(-1,M^2 \tau_{\rm ir}^2)$, with $\Gamma(\alpha,y)$ being the incomplete gamma-function.  Results obtained in the chiral limit are presented in Table\,\ref{Table:Para1}.

\begin{table}[t]
\caption{Results obtained with (in GeV) $m=0$, $m_G=0.11\,$, $\Lambda_{\rm ir} = 0.24\,$, $\Lambda_{\rm uv}=0.823$.  They are commensurate with those from QCD-based DSE studies \protect\cite{Maris:1997tm}.  Dimensioned quantities are listed in GeV or fm, as appropriate, and $\kappa := -\langle \bar q q \rangle^{1/3}$.
\label{Table:Para1}
}
\begin{center}
\begin{tabular*}
{\hsize}
{
c@{\extracolsep{0ptplus1fil}}
c@{\extracolsep{0ptplus1fil}}
c@{\extracolsep{0ptplus1fil}}
c@{\extracolsep{0ptplus1fil}}
|c@{\extracolsep{0ptplus1fil}}
c@{\extracolsep{0ptplus1fil}}
c@{\extracolsep{0ptplus1fil}}
c@{\extracolsep{0ptplus1fil}}
|c@{\extracolsep{0ptplus1fil}}
c@{\extracolsep{0ptplus1fil}}}
${\cal N}$ & $E_\pi^{\rm c}$ & $F_\pi^{\rm c}$ & $F_R$ & M & $\kappa$ &$f_\pi^0$& $\displaystyle\left.f_\pi^0\right|_{F_\pi\to 0}$ & $r_\pi^0$& $\displaystyle\left.r_\pi^0\right|_{F_\pi\to 0}$\\\hline
0.23 & 4.28 & 0.69 & 0.68 & 0.40 & 0.22 & 0.094 & 0.11 & 0.29 & 0.41\\
\end{tabular*} 
\end{center}
\end{table}

Using the interaction we've specified, the homogeneous BSE for the pseudoscalar meson is ($q_+=q+P)$
\begin{equation}
\Gamma_{\pi}(P) = - \frac{4}{3} \frac{1}{m_G^2} \,
\int\frac{d^4 q}{(2\pi)^4} \, \gamma_{\mu} \chi_\pi(q_+,q) \gamma_{\mu}  \,. \label{Gamma-eq}
\end{equation}
With a symmetry-preserving regularisation of the interaction, the Bethe-Salpeter amplitude cannot depend on relative momentum.  Hence Eq.\,(\ref{genpibsa}) can be rewritten
\begin{equation}
\Gamma_\pi(P) = \gamma_5 \left[ i E_\pi(P) + \frac{1}{M} \gamma\cdot P F_\pi(P) \right].
\end{equation}

Preserving the vector and axial-vector Ward-Takahashi identities is crucial when computing properties of the pion.  The $m=0$ axial-vector identity states 
\begin{equation}
\label{avwti}
P_\mu \Gamma_{5\mu}(k_+,k) = S^{-1}(k_+) i \gamma_5 + i \gamma_5 S^{-1}(k)\,,
\end{equation}
where $\Gamma_{5\mu}(k_+,k)$ is the axial-vector vertex, which is determined by
\begin{equation}
\Gamma_{5\mu}(k_+,k) =\gamma_5 \gamma_\mu
- \frac{4}{3}\frac{1}{m_G^2} \int\frac{d^4q}{(2\pi)^4} \, \gamma_\alpha \chi_{5\mu}(q_+,q) \gamma_\alpha\,. \label{aveqn}
\end{equation}
To achieve this, one must implement a regularisation that maintains Eq.\,(\ref{avwti}).  To see what this entails, contract Eq.\,(\ref{aveqn}) with $P_\mu$ and use Eq.\,(\ref{avwti}) within the integrand.  This yields the following two chiral limit identities:
\begin{eqnarray}
\label{Mavwti}
M & = & \frac{8}{3}\frac{M}{m_g^2} \int\! \frac{d^4q}{(2\pi)^4} \left[ \frac{1}{q^2+M^2} +  \frac{1}{q_+^2+M^2}\right],\\
\label{0avwti}
0 & = & \int\! \frac{d^4q}{(2\pi)^4} \left[ \frac{P\cdot q_+}{q_+^2+M^2} -  \frac{P\cdot q}{q^2+M^2}\right]\,,
\end{eqnarray}
which must be satisfied after regularisation.  Analysing the integrands using a Feynman parametrisation, one arrives at the follow identities for $P^2=0=m$:
\begin{eqnarray}
 M & = &  \frac{16}{3}\frac{M}{m_G^2} \int\! \frac{d^4q}{(2\pi)^4} \frac{1}{[q^2+M^2]}, \label{avwtiMc}\\
0 & = & \int\! \frac{d^4q}{(2\pi)^4} \frac{\frac{1}{2} q^2 + M^2 }{[q^2+M^2]^2} \label{avwtiAc}.
\end{eqnarray}

Equation\,(\ref{avwtiMc}) is just the chiral-limit gap equation. Hence it requires nothing new of the regularisation scheme.   On the other hand, Eq.\,(\ref{avwtiAc}) is novel: it states that the axial-vector Ward-Takahashi identity is satisfied if, and only if, the model is regularised so as to ensure there are no quadratic or logarithmic divergences.  Unsurprisingly, these are the just the circumstances under which a shift in integration variables is permitted, an operation required in order to prove Eq.\,(\ref{avwti}).  

One can now write the explicit form of Eq.\,(\ref{Gamma-eq}):
\begin{equation}
\label{bsefinal0}
\left[
\begin{array}{c}
E_\pi(P)\\
F_\pi(P)
\end{array}
\right]
= \frac{1}{3\pi^2 m_G^2}
\left[
\begin{array}{cc}
{\cal K}_{EE} & {\cal K}_{EF} \\
{\cal K}_{FE} & {\cal K}_{FF}
\end{array}\right]
\left[\begin{array}{c}
E_\pi(P)\\
F_\pi(P)
\end{array}
\right],
\end{equation}
where, with $m=0=P^2$, anticipating the Goldstone character of the pion,
%
%
%
%
\begin{equation}
\begin{array}{cl}
{\cal K}_{EE}  =  {\cal C}(M^2;\tau_{\rm ir}^2,\tau_{\rm uv}^2)\,, &  {\cal K}_{EF}  =  0\,,\\
2 {\cal K}_{FE} = {\cal C}_1(M^2;\tau_{\rm ir}^2,\tau_{\rm uv}^2) \,,&
{\cal K}_{FF} = - 2 {\cal K}_{FE}\,.
\end{array}
\end{equation}
Here ${\cal C}_1(z) = - z {\cal C}^\prime(z)$, where we have suppressed the dependence on $\tau_{\rm ir,uv}$.  In order to obtain this form of ${\cal K}_{FF}$, one must employ Eq.\,(\ref{avwtiAc}).  The solution of Eq.\,(\ref{bsefinal0}) gives the pion's chiral-limit Bethe-Salpeter amplitude:
\begin{equation}
\label{EpiFpi1}
E^1_\pi = 0.987 \,,\; F^1_\pi=0.160\,,
\end{equation}
written with unit normalisation.  

The canonical normalisation procedure ensures unit residue for the pion bound-state contribution to the quark-antiquark scattering matrix, a property of
$\Gamma_\pi^{\rm c}(P) =\frac{1}{{\cal N}} \, \Gamma_\pi^1(P)$, where
\begin{equation}
%
{\cal N}^2 P_\mu = N_c\, {\rm tr} \int\! \frac{d^4q}{(2\pi)^4}\Gamma_\pi^1(-P)
 \frac{\partial}{\partial P_\mu} S(q+P) \, \Gamma_\pi^1(P)\, S(q)\,. \label{Ndef}
\end{equation}
In the chiral limit,
\begin{equation}
{\cal N}_0^2 = \frac{N_c}{4\pi^2} \frac{1}{M^2} \, {\cal C}_1(M^2;\tau_{\rm ir}^2,\tau_{\rm uv}^2)
E_\pi^1 [ E_\pi^1 - 2 F_\pi^1].
\label{Norm0}
\end{equation}

The pion's leptonic decay constant is obtained from the canonically normalised amplitude
and in the chiral limit
\begin{equation}
\label{fpi0}
f_\pi^0 = \frac{N_c}{4\pi^2} \frac{1}{M} {\cal C}_1(M^2;\tau_{\rm ir}^2,\tau_{\rm uv}^2)  [ E_\pi^{\rm c} - 2 F_\pi^{\rm c} ]\,.
\end{equation}

If we have preserved Eq.\,(\ref{avwti}), then, in the neighbourhood of $P^2=0$, the solution of Eq.\,(\ref{aveqn}) has the form
\begin{equation}
\Gamma_{5\mu}(k_+,k) = \frac{P_\mu}{P^2} \, 2 f_\pi^0 \, \Gamma^{\rm c}_\pi(P) + \gamma_5\gamma_\mu F_R(P)
\end{equation}
and the following generalised Goldberger-Treiman relations \cite{Maris:1997hd} will hold:
\begin{equation}
\label{GTI}
f_\pi^0 E^{\rm c}_\pi = M \,,\; 2\frac{F^{\rm c}_\pi}{E^{\rm c}_\pi}+ F_R = 1\,.
\end{equation}
That they do can be verified from Table\,\ref{Table:Para1}, which also shows that $F_\pi(P)$, necessarily nonzero in a vector exchange theory, irrespective of the pointwise behaviour of the interaction, has a measurable impact on the value of $f_\pi$, acting here to reduce it by $15$\%.

There is more to be learnt from Eqs.\,(\ref{Norm0}), (\ref{fpi0}).  These chiral-limit expressions exhibit striking similarities.  So, replace $\Gamma_\pi^{\rm c}$ by $\Gamma_\pi^1$ in Eq.\,(\ref{fpi0}) and then take the ratio:
\begin{equation}
\frac{f_\pi^0 {\cal N}_0}{{\cal N}_0^2}= \frac{f_\pi^0}{{\cal N}_0} = \frac{M}{E_\pi^1}\,.
\end{equation}
The last identity is a tautology: it states that $E_\pi^{\rm c} = M/f_\pi^0$, which we already knew from Eq.\,(\ref{avwti}).  This illustrates a general result \cite{Maris:1998hc}:
in a vector exchange theory with DCSB, the canonical normalisation constant for the pion Bethe-Salpeter amplitude is equivalent to its leptonic decay constant when both are evaluated in the chiral limit and a symmetry preserving regularisation procedure is employed.

With the foundation laid, one can evaluate the electromagnetic pion form factor in the generalised impulse approximation; i.e., at leading-order in the symmetry-preserving truncation scheme \cite{Roberts:1994hh,Maris:1998hc,Maris:2000sk}.  Namely, for an incoming pion with momentum $p_1=K-Q/2$, which absorbs a photon with space-like momentum $Q$, so that the outgoing pion has momentum $p_2=K+Q/2$,
\begin{eqnarray}
\nonumber
2 K_{\mu} F_{\pi}^{\rm em}(Q^2) &=& 2 N_c \int\frac{d^4t}{(2\pi)^4}
{\rm tr_D} \Bigg[ i \Gamma_{\pi}^{\rm c}(-p_2) S(t+p_2)\\
&& \times  i \gamma_{\mu}  S(t+p_1) \; i \Gamma_{\pi}^{\rm c}(p_1) \; S(t) \Bigg].
\label{KF}
\end{eqnarray}

The form factor is expressible as a sum; viz.,
\begin{eqnarray}
\label{F123}
\lefteqn{F_{\pi}^{\rm em}(Q^2) = F_{\pi,EE}^{{\rm em}} + F_{\pi,EF}^{{\rm em}} + F_{\pi,FF}^{{\rm em}},}\\
\nonumber & =& T^{\pi}_{EE}(Q^2) E_\pi^{\rm c}\,\!^2 + T^{\pi}_{EF}(Q^2) E_\pi^{\rm c} F_\pi^{\rm c}+ T^{\pi}_{FF}(Q^2) F_\pi^{\rm c}\,\!^2.
\end{eqnarray}
In the chiral limit one finds
\begin{eqnarray}
T^{\pi}_{EE} & = &\frac{N_c}{4\pi^{2}}
\int_{0}^{1} dx \frac{1}{\sigma(x)}\, \mathcal{C}_{1}(\sigma(x))\, , \\
\nonumber
T^{\pi}_{EF} & = & -2 \, T^{\pi}_{EE} \\
& & + \frac{N_c}{4\pi^{2}}
\int_{0}^{1}dx_{1}dx_{2}
\frac{2x_{1}^{2}Q^{2}}{\omega^{2}(x_{1},x_{2})}
 \, \mathcal{C}_{2}(\omega(x_{1},x_{2})) , \label{TpiEF}\\
\nonumber T^{\pi}_{FF}&=&\frac{N_c}{4\pi^{2}}\int_{0}^{1}dx_{1}dx_{2}
\frac{x_{1}^{2}Q^{2}}{M^{2}}  \bigg[\frac{1}{\omega(x_{1},x_{2})} \,
\mathcal{C}_{1}(\omega(x_{1},x_{2}))  \nn \\
&&-
\frac{\omega(x_{1},x_{2})+2M^{2}}{\omega^2(x_{1},x_{2})}\,
\mathcal{C}_{2}(\omega(x_{1},x_{2}))\bigg],  \label{3-components}
\end{eqnarray}
where: ${\cal C}_2(z) = (z^2/2) {\cal C}^{\prime\prime}(z)$ -- we have suppressed the dependence on $\tau_{\rm ir,uv}$; $\sigma(x)=M^{2}+Q^{2}x(1-x)$; and
$\omega(x_{1},x_{2})=M^{2}+Q^{2}x_{1}^{2}x_{2}(1-x_{2})$.  It is straightforward to evaluate the remaining integrals numerically.

\begin{figure}[t] 
\centerline{\includegraphics[clip,width=0.34\textwidth,angle=-90]{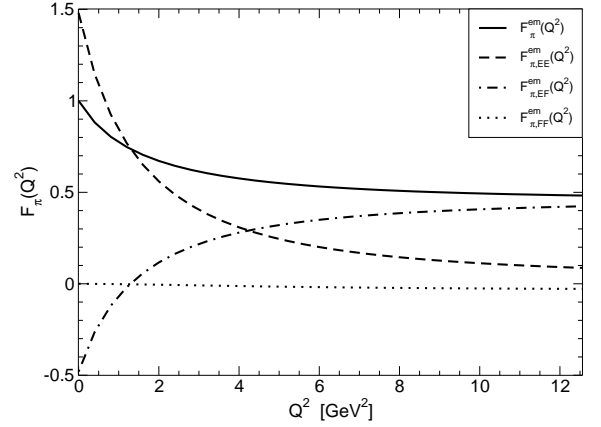}}
\caption{\label{fig1}
$F_{\pi}(Q^2)$ and the separate contributions introduced in Eq.\,(\protect\ref{F123}). 
}
\end{figure}

In Fig.\,\ref{fig1} we plot $F_\pi^{\rm em}(Q^2)$ and the three separate contributions defined in Eq.\,(\ref{F123}).
We highlight two features.  First, $F_{\pi,EF}^{\rm em}(Q^2=0)$ contributes roughly one-third of the pion's unit charge.  This could have been anticipated from Eq.\,(\ref{Norm0}).
Second, and perhaps more dramatic, is that the interaction in Eq.\,(\ref{njlgluon}) generates \begin{equation}
\label{Fpiconstant}
F_\pi^{\rm em}(Q^2 \to\infty) =\,{\rm constant.}
\end{equation}
Both results originate in the nonzero value of $F_\pi(P)$, which is a straightforward consequence of the symmetry-preserving treatment of a vector exchange theory \cite{Maris:1997hd}.  From our perspective, Eq.\,(\ref{Fpiconstant}) is not a surprise: with a symmetry-preserving regularisation of the interaction in Eq.\,(\ref{njlgluon}), the pion's Bethe-Salpeter amplitude cannot depend on the constituent's relative momentum.  This is characteristic of a pointlike particle, which must have a hard form factor.

It is nonetheless worthwhile to elucidate the mathematical origin of Eq.\,(\ref{Fpiconstant}).
As has long been known, $F_{\pi,EE}^{\rm em}$ has its maximum value at $Q^2=0$ and decreases uniformly with increasing $Q^2$: for $Q^2\gg M^2$, $F_{\pi,EE}^{\rm em} \propto M^2/Q^2$.
However, the mixed contribution, $F_{\pi,EF}^{\rm em}$, is significant in magnitude and negative at $Q^2=0$.  As noted above, the sum $F_{\pi,EE}^{\rm em}+F_{\pi,EF}^{\rm em}$ provides a pion with unit charge.  The symmetry-preserving character of our regularisation scheme has ensured this, as can be seen through a comparison of this sum with Eq.\,(\ref{Norm0}).
Very importantly, $F_{\pi,EF}^{\rm em}$ increases with $Q^2$.  It passes through zero at $Q^2 \approx 1.5\,$GeV$^2$ and continues to grow, approaching a nonzero constant value as $Q^2\to \infty$.
These properties are apparent from Eq.\,(\ref{TpiEF}) and follow from the fact that $F_\pi^{\rm c}$ is always multiplied by $\gamma_5\gamma\cdot Q$ in Eq.\,(\ref{KF}).
$F_{\pi,FF}^{\rm em}$ is zero at $Q^2=0$ and evolves to a negative constant as $Q^2\to \infty$.  It is always smaller in magnitude than $F_{\pi,EF}^{\rm em}$.
NB.\ Ref.\,\cite{Maris:1998hc} demonstrated that the presence of pseudovector components alters the asymptotic form of $F_{\pi}^{\rm em}(Q^2)$ by a factor of $Q^2$ cf.\ the result obtained in their absence.

\begin{figure}[t] 
\centerline{\includegraphics[clip,width=0.34\textwidth,angle=-90]{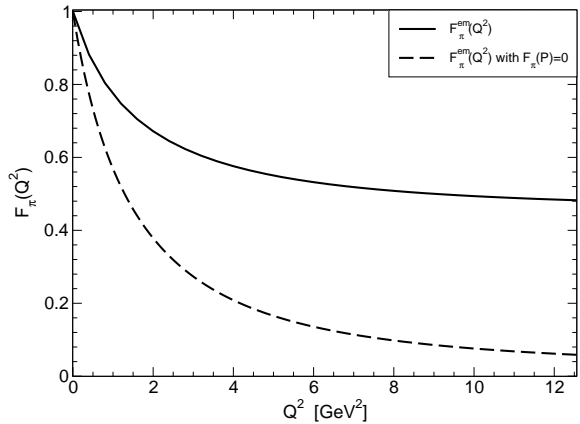}}
\caption{\label{fig2}
$F^{\rm em}_{\pi}(Q^2)$ obtained with the interaction of Eq.\,(\protect\ref{njlgluon}).  \emph{Solid curve} -- complete calculation with the full Bethe-Salpeter kernel; and \emph{dashed curve} -- result obtained with ${\cal K}_{FE}$ artificially set to zero in Eq.\,(\protect\ref{bsefinal0}).}
\end{figure}

A striking feature, evident in Fig.\,\ref{fig1} and emphasised by Fig.\,\ref{fig2},
is that $F_{\pi,EE}^{\rm em}$ is only a good approximation to the net pion form factor for $Q^2 \lesssim M^2$.  $F_{\pi,EE}^{\rm em}$ and $F_{\pi,EF}^{\rm em}$ evolve with equal rapidity -- there is no reason for this to be otherwise, as they are determined by the same mass-scales -- but a nonzero constant comes quickly to dominate over a form factor that falls swiftly to zero.



\begin{figure}[t] 
\centerline{\includegraphics[clip,width=0.34\textwidth,angle=-90]{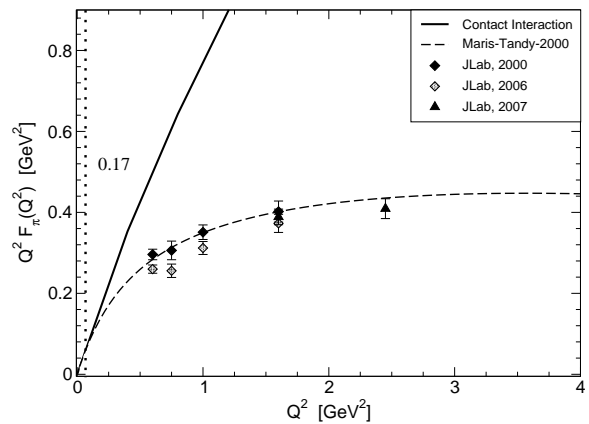}}
\caption{\label{fig4}
\emph{Solid curve} -- $Q^2 F_{\pi}(Q^2)$ obtained with Eq.\,(\protect\ref{njlgluon}).
%
\emph{Dashed curve} -- DSE prediction \protect\cite{Maris:2000sk}, which employed a momentum-dependent renormalisation-group-improved gluon exchange interaction.
For $Q^2>0.17\,$GeV$^2\approx M^2$, marked by the vertical \emph{dotted line}, the contact interaction result for $F_{\pi}(Q^2)$ differs from that in  Ref.\,\protect\cite{Maris:2000sk} by more than 20\%.  The data are from Refs.\,\protect\cite{Volmer:2000ek,Horn:2006tm,Tadevosyan:2007yd}.}
\end{figure}

In Fig.\,\ref{fig4} we compare the form factor computed from Eq.\,(\ref{njlgluon}) with contemporary experimental data \cite{Volmer:2000ek,Horn:2006tm,Tadevosyan:2007yd} and a QCD-based DSE prediction \cite{Maris:2000sk}.  Both the QCD-based result and that obtained from the momentum-independent interaction yield the same values for the pion's static properties. However, for $Q^2>0$ the form factor computed using $\sim 1/k^2$ vector boson exchange is  immediately distinguishable empirically from that produced by a momentum-independent interaction.

Indeed, the figure shows that for $F_\pi^{\rm em}$, existing experiments can already distinguish between different possibilities for the quark-quark interaction.
Treated in precisely the same manner; i.e., as providing the kernel in a symmetry-preserving rainbow-ladder truncation of the DSEs, the contact interaction yields a form factor that is patently in conflict with the data, whereas the result obtained with a momentum-dependent renormalisation-group-improved one-gluon exchange interaction agrees very well.  A dressed-quark propagator and momentum-dependent Bethe-Salpeter amplitude based on this interaction also produces a form factor whose asymptotic power-law behaviour agrees with that of perturbative QCD \cite{Maris:1998hc}, unlike the contact interaction.



\begin{figure}[t] 
\centerline{\includegraphics[clip,width=0.34\textwidth,angle=-90]{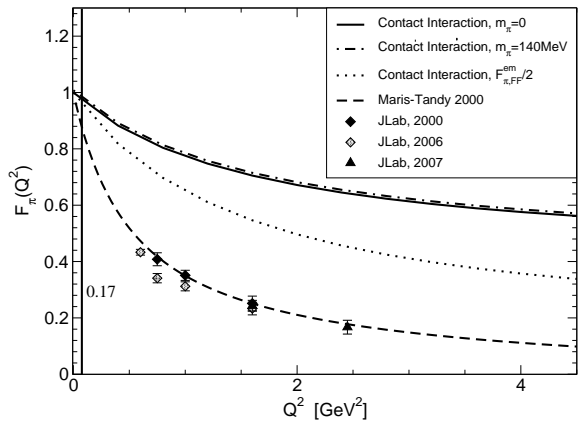}}
\caption{\label{fig5} \emph{Dash-dot curve} -- $F_{\pi}^{\rm em}(Q^2)$ computed with $m_\pi=0.14\,$GeV, yielding $r_\pi=0.290\,$fm; \emph{solid curve} -- chiral limit result, $r_\pi=0.292$; \emph{dotted curve} -- chiral-limit result obtained by suppressing the strength of the pion's pseudovector component by 50\% [see Eq.(\protect\ref{F50})], $r_\pi=0.36\,$fm; and \emph{dashed curve} -- DSE prediction of Ref.\,\protect\cite{Maris:2000sk}, $r_\pi=0.67\,$fm.  The data are from Refs.\,\protect\cite{Volmer:2000ek,Horn:2006tm,Tadevosyan:2007yd}; and, experimentally, $r_\pi=0.672\pm0.008\,$fm \protect\cite{Amsler:2008zzb}.}
\end{figure}

In Fig.\,\ref{fig5} we illustrate two effects.  Firstly, that of a nonzero current-quark mass.  Namely, with $m=8\,$MeV: $m_\pi=0.14\,$, $M=0.41\,$, $f_\pi=0.0945\,$ (in GeV); and
\begin{equation}
E_\pi^1(m_\pi) = 0.987 \,,\; F_\pi^1(m_\pi) = 0.163\,.
\end{equation}
The resulting electromagnetic form factor is depicted in the figure.  Plainly, a nonzero and physical light-quark current-mass has no impact on our discussion.

We also show the impact on $F_\pi^{\rm em}$ of a reduced strength for the pion's pseudovector component.  For this calculation, we suppressed $F_\pi^1(P)$ by 50\%, whilst maintaining unit normalisation; viz., we defined
\begin{equation}
\label{F50}
E_\pi^1=0.997\,,\; F_\pi^1 = 0.080
\end{equation}
cf.\ Eq.\,(\protect\ref{EpiFpi1}), and then recalculated the form factor with the pion Bethe-Salpeter amplitude obtained therefrom.  Notably, this marked suppression of the pseudovector strength has no material impact on our discussion, even though it reduces to just 0.6\% the pseudovector component's contribution to the quadratic norm.   We emphasise that Eq.\,(\ref{F50}) is not a solution of the pion's BSE.  Instead, it is an artificial construction employed solely for the purpose of illustrating the robust nature of our conclusions.

We explored the consequences for pion observables of a momentum-independent vector-exchange interaction, regularised in a symmetry-preserving manner.  With this implementation of the interaction, the results are directly comparable with experiment, computations based on well-defined and systematically-improvable truncations of QCD's DSEs \cite{Maris:2000sk}, and perturbative QCD.  In a vector exchange theory, independent of the pointwise behaviour of the interaction, the pion possesses components of pseudovector origin, which play a material role in the evolution of the pion's form factor away from $Q^2=0$ and, indeed, completely determine the form factor's power-law behaviour at large $Q^2$ \cite{Maris:1998hc}.

We find that the contact interaction, whilst capable of describing pion static properties well, Table\,\ref{Table:Para1}, generates a form factor whose evolution with $Q^2$ deviates markedly from experiment for $Q^2>0.17\,$GeV$^2\approx M^2$ and produces asymptotic power-law behaviour, Eq.\,(\ref{Fpiconstant}), in serious conflict with perturbative-QCD \cite{Farrar:1979aw,Efremov:1979qk,Lepage:1980fj}.

The contact interaction produces a momentum-independent dressed-quark mass function, in contrast to QCD-based DSE studies \cite{Bhagwat:2006tu,Roberts:2007ji} and lattice-QCD \cite{Bowman:2005vx}.  This is fundamentally the origin of the marked discrepancy between the form factor it produces and extant experiment.  Hence our study highlights that form factor observables, measured at an upgraded Jefferson laboratory, e.g., are capable of mapping the running of the dressed-quark mass function.  We are currently working to establish the signals of the running mass in baryon elastic and transition form factors.

We acknowledge constructive input from P.\,C.~Tandy.
This work was supported by:
CIC and CONACyT grants, under project nos.\ 4.10 and 46614-I;
the U.\,S.\ Department of Energy, Office of Nuclear Physics, contract nos.~DE-AC02-06CH11357 and DE-FG03-97ER4014;
and the U.\,S.\ National Science Foundation under grant no.\ PHY-0903991 in conjunction with a CONACyT Mexico-USA collaboration grant.

\bibliography{gbcrpionS}

\end{document}